\documentclass[amsmath,amssymb,aps,twocolumn,pra]{revtex4-2}
\usepackage{tabularx}
\usepackage{graphicx,amsmath,amssymb}
\usepackage{epstopdf}
\usepackage{grffile}
\usepackage[usenames]{color}
\usepackage{float}
\usepackage{color}
\usepackage{mathrsfs}
\begin{document}
\title{Quantum fluctuations and unusual critical exponents in a quantum Rabi Triangle}
\author{Xiao Qin}
\affiliation{Department of Physics, Chongqing Key Laboratory for strongly coupled Physics, Chongqing University, Chongqing 401330, China}
\author{Yu-Yu Zhang}
\email{yuyuzh@cqu.edu.cn}
\affiliation{Department of Physics, Chongqing Key Laboratory for strongly coupled Physics, Chongqing University, Chongqing 401330, China}

\begin{abstract}
Quantum fluctuations of a quantum Rabi triangle are studied using an analytical approach beyond the mean-field theory. By applying an artificial magnetic field among three cavities, time-reversal symmetry breaking is manifested through a directional transfer dynamics of photons. In contrast to previous studies, we focus on the scaling exponents of the fluctuations of the local photon number and the position variance near the critical point. By accurate calculation using Bogoliubov transformation we show that two scaling laws emerge respectively for the frustrated cavity and the remaining cavities, which are associated with the geometric frustrations. Especially, for the frustrated cavity, the scaling exponent in the chiral superradiant phase is different from that in the frustrated antiferromagnetic superradiant phase without an artificial magnetic field. The unusual scaling exponents predict distinct universality classes from the single-cavity Rabi universality.
We suggest that the accurate critical exponents in few-body system is useful for identifying exotic quantum phase transition in light-matter coupling system.    
\end{abstract}

\date{\today }
\maketitle
\section{Introduction}
The interaction between light and matter has brought forth a new class of quantum many-body system in understanding of strongly correlated systems and quantum phase transition ~\cite{NaturePhysics2006,PhysRevLett2007,PhysRevLett103,PhysRevLett109}. The effect of quantum fluctuations driving the quantum phase transitions is especially pronounced in characterizing singularity and universality classes by universal scaling laws~\cite{voita2003,sachdev2011,scaling1996,RevModPhys1997}. A superradiant phase transition (SPT) is a remarkable phenomenon in the Dicke mode~\cite{PhysRev1954,Emary03,PhysRevLett2011,PhysRevA2008,PhysRevA2009}, which describes an ensemble of two-level atoms interacting with a quantized single-mode cavity. Such SPT has been found in few-body systems such as the quantum Rabi model in the limit where the atomic transition frequency in the unit of the cavity frequency tends to infinity~\cite{hwang2015,Ashhab2013,liu2017,chen2020,chen2021,NCcai2021}, exhibting the same critical exponents as the Dicke model~\cite{hwang2015,PhysRevA2008,PhysRevA2009}. A few-body system of light-atom interactions sheds new light in quantum simulation of quantum phase transitions due to its high control and tunability.

To explore intriguing quantum phases, synthetic magnetic fields are applied in atom-cavity coupling systems and photon lattices~\cite{dalibard2011}, chiral edge currents in atoms~\cite{PhysRevLett122,PhysRevLett124}, chiral ground-state currents of interacting photons~\cite{roushan2017}, and fractional quantum Hall physics in the Jaynes-Cummings Hubbard lattice~\cite{hayward2012,hayward2016}. 
In the presence of an artificial magnetic field, unusual superradiant phases have been found in the generalized Rabi and Dicke systems, including a chiral phase in a quantum Rabi triangle ~\cite{PhysRevLettzhang2021}, chiral superradiant phases analogies to quantum magnetism in a quantum Rabi ring~\cite{PhysRevLettzhang2022} and anomalous superradiant phases in a Dicke lattice model~\cite{zhao2022anomalous}. Such exotic SPTs exhibit unusual scaling behaviors by comparing to the conventional SPT in the Dicke model. Besides the artifical magnetic field, geometric frustration has been proved to induce counterintuitive critical phenomena. Different critical exponents of the excitation energy have been found on the
two sides of phase transitions in the Rabi triangle  ~\cite{PhysRevLettzhang2022} and in the Dicke trimer ~\cite{PRLHuang2022,zhao2022anomalous}, which are associated the frustrated geometry. However, there appears to be no clear scaling exponents of quantum fluctuations near the critical value.  Since there is a challenge of an accurate solution beyond a mean-field approach for the Rabi and Dicke lattice including interactions between cavities. 

In this paper, we perform an analytical solution beyond the mean-field approximation for the quantum Rabi triangle with an artificial magnetic filed. In contrast to previous studies, we present an analytical expression of quantum fluctuations and obtain accurate scaling exponents near the critical point. In the chiral superradiant phase (CSP), we find the fluctuation of the mean photons in the frustrated cavity diverges with an anomalous critical exponent, which is distinct from the conventional exponent of the remaining cavities.  In the absence of an artificial magnetic field, the scaling exponent of the frustrated cavity in the frustrated antiferromagnetic superradiant phase (AFSP) is different from that in the CSP.
Thus, the scaling behavior of quantum fluctuations in the CSP and the frustrated AFSP fall into two classes, one for the frustrated cavity and the other for the remaining cavities. Moreover, the scaling behavior below and above the transition point is different due to the frustrated local photons. In contrast, the scaling exponents are equal to each other on the two sides of the transition of the ferromagnetic superradiant phase (FSP), which are the same as that in the Dicke model.  

The rest of this paper is organized as following: In Sec. II,
we introduce the Hamiltonian of the quantum Rabi triangle and the dynamics of photons with an artificial magnetic field. In
Sec. III, the quantum fluctuations of the mean photons and the position variance are given analytically in the normal phase below the critical value.  
Sec. IV gives the analytical solutions for different superradiant phases and the expressions for the quantum fluctuations. Sec. V gives the scaling exponents of the quantum fluctuations and excitation energies beyond the mean-field approximation. The conclusion is given in Sec. VI.

\section{Hamiltonian}
A Rabi triangle model consists of $N=3$ coupled cavities, and the Hamiltonian is
\begin{equation}\label{Ham}
H_{\mathtt{QRT}}=\sum_{n}^{3}H_{R,n}+\sum_{n}^{3}J(e^{i\mathcal{\theta }%
}a_{n}^{\dagger }a_{n+1}+e^{-i\mathcal{\theta }}a_{n}a_{n+1}^{\dagger }),
\end{equation}%
where the quantum Rabi Hamiltonian of each cavity coupled to a two-level atom is desribed as $H_{R,n}=\omega a_{n}^{\dagger
}a_{n}+g\left( a_{n}^{\dagger }+a_{n}\right) \sigma _{n}^{x}+\frac{\Delta }{2%
}\sigma _{n}^{x}$. The second term of the Hamiltonian $H_{\mathtt{QRT}}$
describes the photons hopping between the nearest-neighbor cavities with a
phase $\theta $, which can be realized on cavities by dynamical
modulating the cavity-atom coupling and photon hopping strengths~\cite{PhysRevLettzhang2021}. An artificial vector potential $A(r)$ leads the photon hopping terms between nearby cavities $n$ and $m$ to become complex with the
phase given by $\theta =\int_{r_{n}}^{r_{m}}$ $A(r)dr$. The time-reversal symmetry (TRS) is artificially broken when $\theta \neq m\pi \,(m\in \mathbb{Z})$.  
The effective magnetic flux is $3\theta$ in three cavities, which form a closed lopp. 

To explore the effects of the artificial magnetic field we study the dynamics of photon flow in our system.
At $t=0$, a photon is prepared in the $1$st cavity and the two-level atom in each cavity is in the down state, giving the initial state $|\varphi(0)\rangle=|100\rangle|\downarrow\downarrow\downarrow\rangle$. The time-evolving wave function is 
$|\varphi (t)\rangle =e^{-iH_{\mathtt{QRT}}t}|\phi_0\rangle|\varphi$. The mean photon in the $i$-th cavity is given by $N_{i}=\langle \varphi (t)| a_{i}^{\dagger }a_{i}|\varphi (t)\rangle$ .

\begin{figure}[tbp]
\includegraphics[trim=100 100 200 20,scale=1.2]{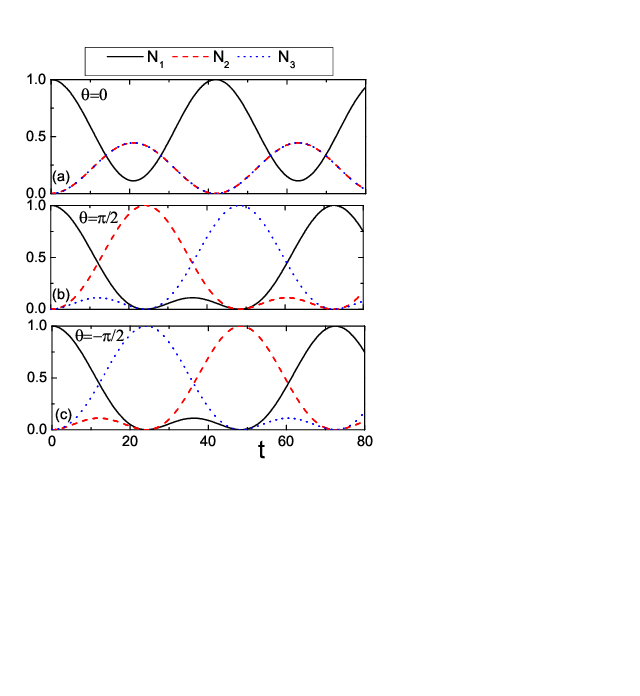}
\caption{Mean photon number in each cavity $N_1$ (black solid line), $N_2$ (red dashed line), $N_3$ (blue dotted line) as a function of time $t$ for three values of (a)$\theta=0$, (b)$\pi/2$ and (c)$-\pi/2$ for the scaled coupling strength $g_1=0.1$. In this paper we use $\Delta/\omega=50$ and $J/\omega=0.05$ by choosing $\omega=1$ as the unit for frequency.} \label{dynamics}
\end{figure}

Fig.\ref{dynamics} shows the mean photon number evolution in each cavity dependent on $\theta$. The photon number is initially occupied in the first cavity with $N_1=1$.  For $\theta=0$ in Fig.\ref{dynamics}(a), the photon transfer symmetrically from cavity $1$ to cavity $2$ and cavity $3$ simultaneously, then back to cavity $1$. There is no preferred circulation direction. Since TRS is preserved for the trivial case $\theta=0$, the states at time $t$ and $T-t$ satisfy  $\varphi(t)=\varphi(T-t)$ with a time-evolving period $T$.  A completely different dynamics is observed for  $\theta=\pi/2$ in Fig.\ref{dynamics}(b). The photon is flowing unidirectionally, first from cavity $1$, to cavity $2$, to cavity $3$, eventually back to cavity $1$. Such chiral current direction is a signature of the breaking of TRS. Because the evolution of the state from $t=T$ backwards is different as going forwards from $t=0$. Choosing $\theta=-\pi/2$ leads to the opposite direction of the chiral photon flow in Fig.\ref{dynamics}(c). It demonstrate that the artificial flux $\theta$ leads to the breaking of TRS, which behaves similarly to a magnetic flux. 

Besides the TRS breaking induced by the artificial magnetic flux, rich superradiant phase transitions have been explored dependent on $\theta$ in the quantum Rabi triangle ~\cite{PhysRevLettzhang2021,PhysRevLettzhang2022}. The phase transitions occur in the limit where $\Delta$ is much larger than frequency scales in the system $\Delta/\omega\rightarrow\infty$. 
Different from previous studies, we extract unusual critical exponents to classify the universality classes of the nontrivial phase transition by involving the geometric frustrations and the effects of the magnetic flux. In particular, quantum fluctuations are crucial to capture the singularity and scaling exponents near the critical point using an accurate solution beyond a mean-field approximation. 

\section{Quantum fluctuations in the normal phase}
We perform a Schrieffer-Woff transformation with an unitary operator $S_{n}=\exp [-i\sigma _{n}^{y}g_{1}\sqrt{\omega /\Delta }\left(
a_{n}^{\dagger }+a_{n}\right) ]$. The lower-energy Hamiltonian is obtained
by projecting $H_{\mathtt{iCP}}$ to the spin subspace $|\downarrow \rangle $%
, giving
\begin{eqnarray}
H_{\mathtt{NP}}^{\downarrow }&=&\sum_{n=1}^{3}(\omega -2\omega
g_{1}^{2})a_{n}^{\dagger }a_{n}-\omega g_{1}^{2}(a_{n}^{2}+a_{n}^{\dagger
2})\nonumber\\
&&+J\sum_{<nn^{\prime }>}^{3}(e^{i\theta }a_{n}^{\dagger }a_{n^{\prime
}}+h.c.)+E_{0},  \label{icp1}
\end{eqnarray}%
where the energy constant is $E_{0}=3[-\Delta /2-\omega g_{1}^{2}+(\omega
+J)\omega ^{2}g_{1}^{2}/\Delta ]$.

By introducing the Fourier transformation $a_{n}^{\dagger
}=\sum_{q}e^{inq}a_{q}^{\dagger }/\sqrt{N}$ with the quasi-momentum $q=0,\pm
2\pi /3$, the transformed Hamiltonian is expressed
as $H_{\mathtt{iCP}}^{\downarrow }=\sum_{q}\omega _{q}a_{q}^{\dagger
}a_{q}-\omega g_{1}^{2}(a_{q}a_{-q}+a_{q}^{\dagger }a_{-q}^{\dagger })+E_{0}$%
, where $\omega _{q}=\omega -2\omega g_{1}^{2}+2J\cos (\theta -q)$. By
performing a unitary transformation $S_{q}=\exp [\lambda _{q}(a_{q}^{\dagger
}a_{-q}^{\dagger }-a_{q}a_{-q})]$ with a variational squeezing parameter $%
\lambda _{q}=-\frac{1}{8}\ln \frac{\omega _{q}+\omega _{-q}-4\omega g_{1}^{2}%
}{\omega _{q}+\omega _{-q}+4\omega g_{1}^{2}}$, we derive the Hamiltonian in
diagonalized form as $H_{\mathtt{iCP}}^{\downarrow }=\sum_{q}\varepsilon
_{q}a_{q}^{\dagger }a_{q}+E_{g}$ with the
ground-state energy $E_{g}=E_{0}+\frac{1}{2}\sum_{q}(\varepsilon
_{q}-\omega _{q})$. The excitation spectrum is obtained as 
\begin{equation}
\varepsilon _{q}^{\mathtt{NP}}=\frac{1}{2}(\sqrt{\Omega _{+,q}^{2}-16\omega
^{2}g_{1}^{4}}+\Omega _{-,q})
\end{equation}%
with the dispersion $\Omega _{\pm,q}=\omega _{q}\pm\omega _{-q}$. The lowest excitation energy is associated with the qusi-momentum $q$ dependent on $\theta$. For $\theta=0$, the excitation energy $\varepsilon _{q=\pm2/3}^{\mathtt{NP}}$ with $q=\pm2\pi/3$ becomes the lowest one, and decreases to zero for $g_1$ approaching $g_c$ from below in Fig.~\ref{excitation}(a). As $\theta$ increases, the lowest energy is determined by $q=-2\pi/3$ in Fig.~\ref{excitation}(b). Then it changes to $\varepsilon _{q=0}^{\mathtt{NP}}$ with $q=0$ in Fig.~\ref{excitation}(d). The vanishing of $\varepsilon _{q}^{\mathtt{NP}}=0$ dependent
on $\theta $ and $q$-momentum gives the critical
scaled coupling strength $g_{1c}(q)=\frac{1}{2}\sqrt{\frac{1+4J/\omega \cos
\theta \cos q+4J^{2}/\omega ^{2}\cos (\theta +q)\cos (\theta -q)}{%
1+2J/\omega \cos \theta \cos q}}$, which gives the phase boundary. Especially for $\theta=0$, the critical coupling strength is $g_c(\pm2\pi/3)$ originating from the lowest excitation energy with $q=\pm2\pi/3$. By increasing the magnetic flux, the critical flux is given by $\theta_c=\pm cos^{-1}[-2J/(\sqrt{8J^2+\omega^2}+\omega)]$ by solving $g_c(\pm2\pi/3)=g_c(0)$,  which classifies phase boundary by tuning $\theta$.

The ground-state in the NP is 
\begin{equation}
|\varphi _{np}\rangle =\prod_{q}e^{\lambda _{q}(a_{q}^{\dagger
}a_{-q}^{\dagger }-a_{q}a_{-q})}|0\rangle_q
|\downarrow \rangle,
\end{equation}
where $|\downarrow \rangle $ is the lowest state of the atom. The
average photon number in the ground state can be obtained as $\langle
a_{q}^{\dagger }a_{q}\rangle _{np}=\langle \varphi _{np}|a_{q}^{\dagger
}a_{q}|\varphi _{np}\rangle =(\cosh 4\lambda _{q}-1)/2$, which desribes the quantum fluctuation of photons for $g_1$ approaching the critical value $g_{1c}$ from below. The local mean
photon number in the $n$th cavity is obtained by
\begin{equation}\label{local photon}
\langle a_{n}^{\dagger }a_{n}\rangle _{np}=\frac{1}{N}\sum_{q}\langle
a_{q}^{\dagger }a_{q}\rangle _{np}=\frac{1}{2N}\sum_{q}[\frac{\Omega _{+,q}}{2\varepsilon _{q}-\Omega _{-,q}}-1].
\end{equation}
The variance of $x_{q}=a_{q}+a_{q}^{\dagger }$ and $p_{q}=i(a_{q}^{\dagger
}-a_{q})$ are derived as $(\Delta x_{q})^{2}=\langle \varphi _{np}| x_{q}^{2}|\varphi _{np}\rangle -\langle \varphi _{np}|x_{q}|\varphi _{np}\rangle
^{2}=e^{4\lambda _{q}}$, and $(\Delta p_{q})^{2}=\langle p_{q}^{2}\rangle -\langle p_{q}\rangle
^{2}=e^{-4\lambda _{q}}$. The local variance of quadratures in the $n$th cavity is
\begin{eqnarray}\label{position}
(\Delta x_{n})^{2}=\frac{1}{N}\sum_{q}(\Delta x_{q})^{2}=\frac{\Omega _{+,q}+4\omega g_{1}^{2}}{2\varepsilon _{q}-\Omega _{-,q}},
\end{eqnarray}%
and $(\Delta p_{n})^{2}=\frac{1}{N}\sum_{q}e^{-4\lambda _{q}}$.

It indicates that the singularity of the mean photon number $\langle
a_{n}^{\dagger }a_{n}\rangle _{np}$, the variance of $(\Delta x_{n})^{2}$ and $(\Delta p_{n})^{2}$ is
determined by the excitation energy $\varepsilon _{q}$ and $\Omega _{-,q}$ in the denominator.
Obviously, $\varepsilon _{q}$ becomes zero at the critical value $g_{1c}$.
It is interesting to understand the divergence from $\Omega_{-,q}=4J\texttt{sin}\theta \texttt{sin}q$ dependent on $q$ and $\theta$.
In the absence of the magnetic field with $\theta=0$ or the momentum $q=0$, one has $\Omega _{-,q}=0$. It leads to the singularity of quantum fluctuations of $\langle a_{n}^{\dagger }a_{n}\rangle _{np}$ at the critical point in Fig.~\ref{photons}(a) (d) as well as $(\Delta x_{n})^{2}$ in Fig.~\ref{fluctuation}(a)(d). In contrast, for $\theta\neq0$ and  $q=\pm2\pi/3$, it gives nonzero value of $\Omega _{-,q}$, which results in a finite value of $\langle a_{n}^{\dagger }a_{n}\rangle _{\texttt{NP}}$ and $(\Delta x_n)^{2}_{\texttt{NP}}$, respectively. There appear non-divergent fluctuations in Fig.~\ref{photons}(b) and Fig.~\ref{fluctuation}(b) for $g_1$ approaching $g_{1c}$ from below.

\begin{figure}[tbp]
\includegraphics[scale=0.18]{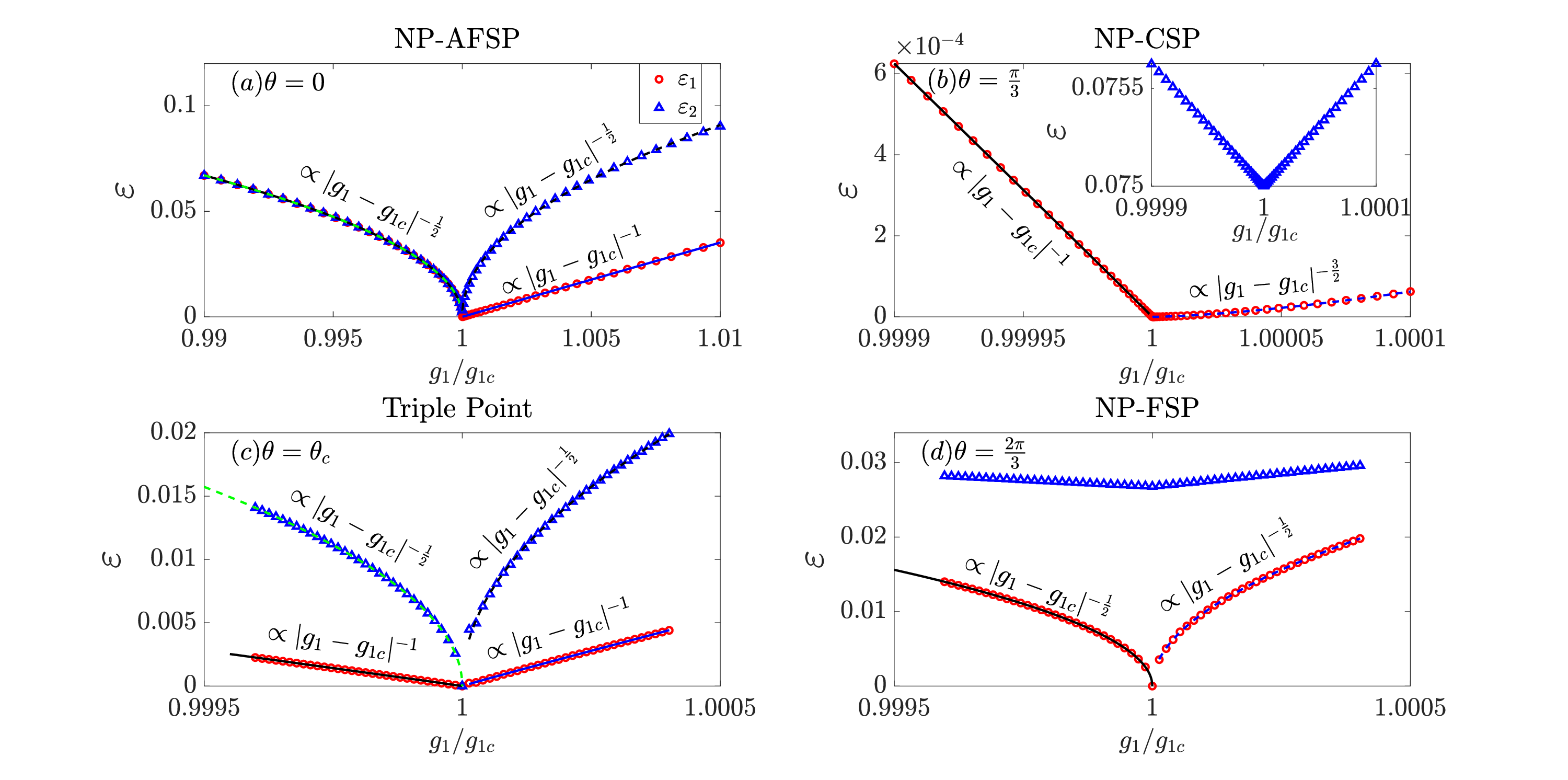}
\caption{The lowest and second excitation energy $\varepsilon_{1(2)}$ below and above the critical coupling strength $g_{1c}$ for the NP-AFSP transition ($\theta=0$)(a), the NP-CSR transition ($\theta=0.1<\theta_c$)(b),
the triple point ($\theta=\theta_c$)(c) and the NP-FSP transition ($\theta=1.7>\theta_c$) (d). 
The parameters are $\Delta/\omega=100$ and $J/\omega=0.05$ by setting $\omega=1$ as the unit for the frequency. The inset (b) shows the second excitation energy $\varepsilon_{2}$ for the NP-CSP transition, which exhibits an obvious energy gap in comparison to $\varepsilon_{1}$. }
\label{excitation}
\end{figure}

\section{Quantum fluctuations in the superadiant phase}
 It is interesting to explore the quantum fluctuations and scaling exponents when $g_1$ approaches $g_{1c}$ from above. As the atom-cavity coupling increases $g_1>g_{1c}$, rich superradiant phases emerge by adjusting the flux $\theta$~\cite{PhysRevLettzhang2022,PhysRevLettzhang2021}. Since the photon population in each cavity becomes macroscopic. 
The bosonic operator $a_{n}^{\dagger }$ $\left( a_{n}\right) $ is expected
to be shifted as $\tilde{a}_{n}=D^{\dagger }(\alpha _{n})a_{n}D(\alpha _{n})=a_{n}+\alpha
_{n}$ with the complex displacement $\alpha _{n}=A_{n}+iB_{n}$. It is different from the mean field approximation, for which the bosonic operator $a_n$ is replaced by its mean value $\langle a_{n}\rangle=\alpha _{n}$. The Hamiltonian in the superradiant phases becomes 
\begin{eqnarray}
H_{\text{SR}}^{\downarrow }&=&\sum_{n=1}^{3}\omega \tilde{a}_{n}^{\dagger }a_{n}-%
\frac{\Delta _{n}}{2}\tau_n^z+\lambda_n\left( \tilde{a}_{n}^{\dagger }+\tilde{a}_{n}\right)\tau_n^x
\nonumber\\
&&+Ja_{n}^{\dagger }(e^{i\theta }\tilde{a}_{n+1}+e^{-i\theta }\tilde{a}_{n-1})+V_{\text{off}}+E_0,
\label{SRhm}
\end{eqnarray}
where the transformed Pauli matrix is $\tau_n^z=\Delta/\Delta_n\sigma_n^z+4gA_n/\Delta_n\sigma_n^x$ with the renormalized parameter $\Delta_n=\sqrt{\Delta^2+16g^2A_n^2}$, and the effective coupling strength is $\lambda_n=g\Delta/\Delta_n$. The off-diagonal term is expressed as $V_{\text{off}}=\sum_{n}\omega (\alpha _{n}a_{i}^{\dagger }+\alpha
_{n}^{\ast }a_{n})+g\left( a_{n}^{\dagger }+a_{n}\right) \sin (2\gamma
_{n})\sigma _n^{z}+J[a_{n}^{\dagger }(e^{i\theta }\alpha _{n+1}+e^{-i\theta
}\alpha _{n-1})+h.c.]$. By eliminating $V_{\text{off}}$ term, we obtain equations explicitly
\begin{eqnarray}\label{real} 
0&=&\omega A_{n}-g\sin (2\gamma _{n})+J[(A_{n+1}+A_{n-1})\cos \theta \nonumber\\
&+&(B_{n-1}-B_{n+1})\sin \theta ]=0,  
\end{eqnarray}
and 
\begin{equation}
0=\omega B_{n}+J[\cos \theta (B_{n+1}+B_{n-1})+\sin \theta(A_{n+1}-A_{n-1})].  
\end{equation}
$\alpha_n$ can be accurately obtained by solving above equations.
The lowest-energy Hamiltonian is obtained by projecting to the spin subspace $
|\downarrow \rangle $, giving
\begin{eqnarray}
H_{\text{eff}}^{\downarrow }&=&\sum_{n=1}^{3}\omega a_{n}^{\dagger }a_{n}-%
\frac{\lambda _{n}^{2}}{\Delta _{n}}\left( a_{n}^{\dagger }+a_{n}\right)
^{2}\nonumber\\
&&+Ja_{n}^{\dagger }(e^{i\theta }a_{n+1}+e^{-i\theta }a_{n-1})+E_{g}.
\label{hmccp2}
\end{eqnarray}%
where the ground-state energy is $E_{g}=\sum_{n}\omega \alpha _{n}^{\ast }\alpha _{n}+J\sum_{n}\alpha
_{n}^{\ast }(e^{i\theta }\alpha _{n+1}+e^{-i\theta }\alpha _{n-1})-\Delta _{n}/2$.

Superradiant phases can be characterized by the order parameter $\alpha_n$. Firstly, for $\theta=0$ and $g_1>g_{1c}^{\text{AFSP}}(\pm2\pi/3)$, the system is in the frustrated AFSP. $\alpha _{n}$ is real with opposite signs for the neighboring cavities. It corresponds to an AFSP with antiferromagnetic order. Moreover, geometric frustration in three cavities yields site-dependent $\alpha _{n}$, so-called frustrated AFSP. The ground state breaks the $C_3$ symmetry. Secondly, for $0<\theta<\theta_c$, the system enters the CSP regime above the critical coupling strength $g_{c}^{\text{CSP}}(-2\pi/3)$~\cite{PhysRevLettzhang2021}. $\alpha_n$ is complex and $n$ dependent, which is different from that in the AFSP. For $\theta_c<\theta\leq\pi$ the displacement in each cavity $\alpha_n$ is real and is independent of $n$, which gives $\alpha _{n}=\pm\sqrt{16g^{4}/(\omega+2Jcos\theta)^2-\Delta^2}/(4g)$. It forms a ferromagnetic order with the same displacement of the neighboring cavities for  $g_1>g_{1c}^{\text{FSP}}(q=0)$. It corresponds to the FSP.

The Hamiltonian in Eq. (\ref{hmccp2}) is bilinear in the creation and
annihilation operators $a_{i}^{\dagger }$ and $a_{i}$, which captures the quantum fluctuations especially near the critical coupling strength $g_c$. It can be
diagonalized by the bosonic Bogoliubov transformation. By using the
denotation $\alpha =\{a_{1},a_{2},a_{3},a_{1}^{\dagger },a_{2}^{\dagger
},a_{3}^{\dagger }\}$, the Hamiltonian in Eq. (\ref{hmccp2}) can be written
in matrix form as $H_{\text{eff}}^{\downarrow }=\alpha M\alpha ^{\dagger
}-3(\omega -\lambda _{n}^{2}/\Delta _{n})/2$, where a transformed matrix $M$
is
\begin{widetext}
\begin{equation}
M=\left(
\begin{array}{cccccc}
\omega /2-\lambda _{1}^{2}/\Delta _{1} & Je^{-i\theta }/2 & Je^{i\theta }/2
& -\lambda _{1}^{2}/\Delta _{1} & 0 & 0 \\
Je^{i\theta }/2 & \omega /2-\lambda _{2}^{2}/\Delta _{2} & Je^{-i\theta }/2
& 0 & -\lambda _{2}^{2}/\Delta _{2} & 0 \\
Je^{-i\theta }/2 & Je^{i\theta }/2 & \omega /2-\lambda _{3}^{2}/\Delta _{3}
& 0 & 0 & -\lambda _{3}^{2}/\Delta _{3} \\
-\lambda _{1}^{2}/\Delta _{1} & 0 & 0 & \omega /2-\lambda _{1}^{2}/\Delta
_{1} & Je^{i\theta }/2 & Je^{-i\theta }/2 \\
0 & -\lambda _{2}^{2}/\Delta _{2} & 0 & Je^{-i\theta }/2 & \omega /2-\lambda
_{2}^{2}/\Delta _{2} & Je^{i\theta }/2 \\
0 & 0 & -\lambda _{3}^{2}/\Delta _{3} & Je^{i\theta }/2 & Je^{-i\theta }/2 &
\omega /2-\lambda _{3}^{2}/\Delta _{3}%
\end{array}%
\right) .  \label{Mmatrix}
\end{equation}
\end{widetext}

We perform a Bogoliubov's transformation to give bosonic
operators $\beta =\{b_{1}^{\dagger },b_{2}^{\dagger },b_{3}^{\dagger
},b_{1},b_{2},b_{3}\}$ as a linear combination of $\alpha =\{a_{1}^{\dagger
},a_{2}^{\dagger },a_{3}^{\dagger },a_{1},a_{2},a_{3}\}$, which satisfies $%
\alpha ^{\dagger }=T\beta ^{\dagger }$ with a paraunitary matrix $T$ rather
than a unitary matrix. To ensure the bosonic commutation relation, the
paraunitary matrix $T$ satisfies the relations of $T^{\dagger }\Lambda
T=T\Lambda T^{\dagger }=\Lambda  $, where $\Lambda =$ $\left(
\begin{array}{cc}
I_{3\times3} & 0 \\
0 & -I_{3\times3}%
\end{array}%
\right) $ with $I_{3\times3}$ being the identity matrix of order $3$. Substituting for $\alpha $
and $\alpha ^{\dagger }$ in terms of $\beta ^{\dagger }$ and $\beta $, one
obtains the diagonalized form as
\begin{equation}\label{diagham}
H_{\text{eff}}^{\downarrow }=\beta T^{\dagger }MT\beta ^{\dagger
}=2\sum_{k=1}^{3}\varepsilon _{k}b_{k}^{\dagger }b_{k}+\frac{\varepsilon
_{k}-\omega }{2},
\end{equation}%
where $\varepsilon _{k}$ is the corresponding eigenvalue for each excitation mode. The eigenvalues $\pm \varepsilon _{n}$ are obtained by
diagonalizing the matrix $\Lambda M$ as $T^{-1}\Lambda MT=\Lambda
\varepsilon $. The corresponding eigenvector gives the $k$-th column vector of the
paraunitary matrix as $T_{k}=[T_{1k},T_{2k},\cdots ,T_{6k}]^{T}$.

The ground state in the superradiant phases is the vacuum state 
\begin{equation}
|\varphi_{\texttt{SR}}\rangle=|0\rangle_{b_1}|0\rangle_{b_2} |0\rangle_{b_3},    
\end{equation}
where $|0\rangle_{b_n}$ satisfies $b_n|0\rangle_{b_n}=0$. Note that the operator $b_n$ corresponds to the original operator $a_n$ with the displacement transformation $D(\alpha)$ and the transformation $\alpha ^{\dagger }=T\beta ^{\dagger }$. One obtains $a_{n}=\sum_{i=1}^{3}T_{n,i}b_{i}+T_{n,i+3}b_{i}^{\dagger}+\alpha_n$. The
local photon number in the $n$th cavity of the ground state $|\varphi_{\texttt{SR}}\rangle$ is derived explicitly as 
\begin{eqnarray}\label{aa}
\langle a_{n}^{\dagger }a_{n}\rangle
=|T_{n4}|^{2}+|T_{n5}|^{2}+|T_{n6}|^{2}+|\alpha _{n}|^{2}.
\end{eqnarray}
 The variance of $x_{n}$ and $p_{n}$ are
\begin{equation}\label{xp}
(\Delta x_{n})^{2}=|T_{n1}+T_{n4}^{\ast }|^{2}+|T_{n2}+T_{n5}^{\ast
}|^{2}+|T_{n3}+T_{n6}^{\ast }|^{2},
\end{equation}%
and $(\Delta p_{n})^{2}=|T_{n4}^{\ast }-T_{n1}|^{2}+|T_{n5}^{\ast}-T_{n2}|^{2}+|T_{n6}^{\ast }-T_{n3}|^{2}$,
which are derived in detail in Appendix. Thus the quantum fluctuations of $(\Delta x_{n})^{2}$ and $\langle a_{n}^{\dagger }a_{n}\rangle$ are obtained beyond the mean-field approaximation. We calculate the scaling behaviors of the critical fluctuations in the vicinity of the critical value $g_c$ in the following.

\section{Unusual critical exponents}
Our major interests are the scaling exponents of the quantum fluctuations as $g_1$ approaches $g_c$ for the second-order phase transitions from the NP to different superradiant phases.
Generally, the energy gap which is measured by the lowest excitation energy $\varepsilon _{1}$ vanishes as $g_1$ approaches $g_{1c}$
\begin{equation}\label{vanishenergy}
 \varepsilon _{1}\propto|g_1-g_{1c}|^{\gamma}.   
\end{equation}
Here the critical exponent $\gamma=z\nu$ is usually universal~\cite{sachdev2011}, which is independent of most of the parameters of the Hamiltonian. In addition to the vanishing energy gap, a length scale which measures the correlations at the longest distances becomes diverge at the critical value. The variance of position quadrature plays an analogous role of the diverging length scale, which diverges as 
\begin{equation}
 \Delta x\propto|g_1-g_{1c}|^{-\nu}
\end{equation}
with a critical exponent $\nu$. It leads to the dynamical critical exponent $z=\gamma/\nu$. Meanwhile, the fluctuations of the local mean photons near the critical value diverges as 
\begin{equation}
 \langle a_{n}^{\dagger}a_{n}\rangle \propto|g_1-g_{1c}|^{-\beta},   
\end{equation}
where $\beta$ is a critical exponent.
Using the Bogoliubov's diagonalization method, we calculate the scaling exponents of the lowest excitation energy $\varepsilon _{1}$, the fluctuation of position variance $(\Delta x)^2$ in Eq.(\ref{xp}) and photons $\langle a_{n}^{\dagger}a_{n}\rangle$ in Eq.(\ref{aa}). 

\begin{figure}[tbp]
\includegraphics[scale=0.11]{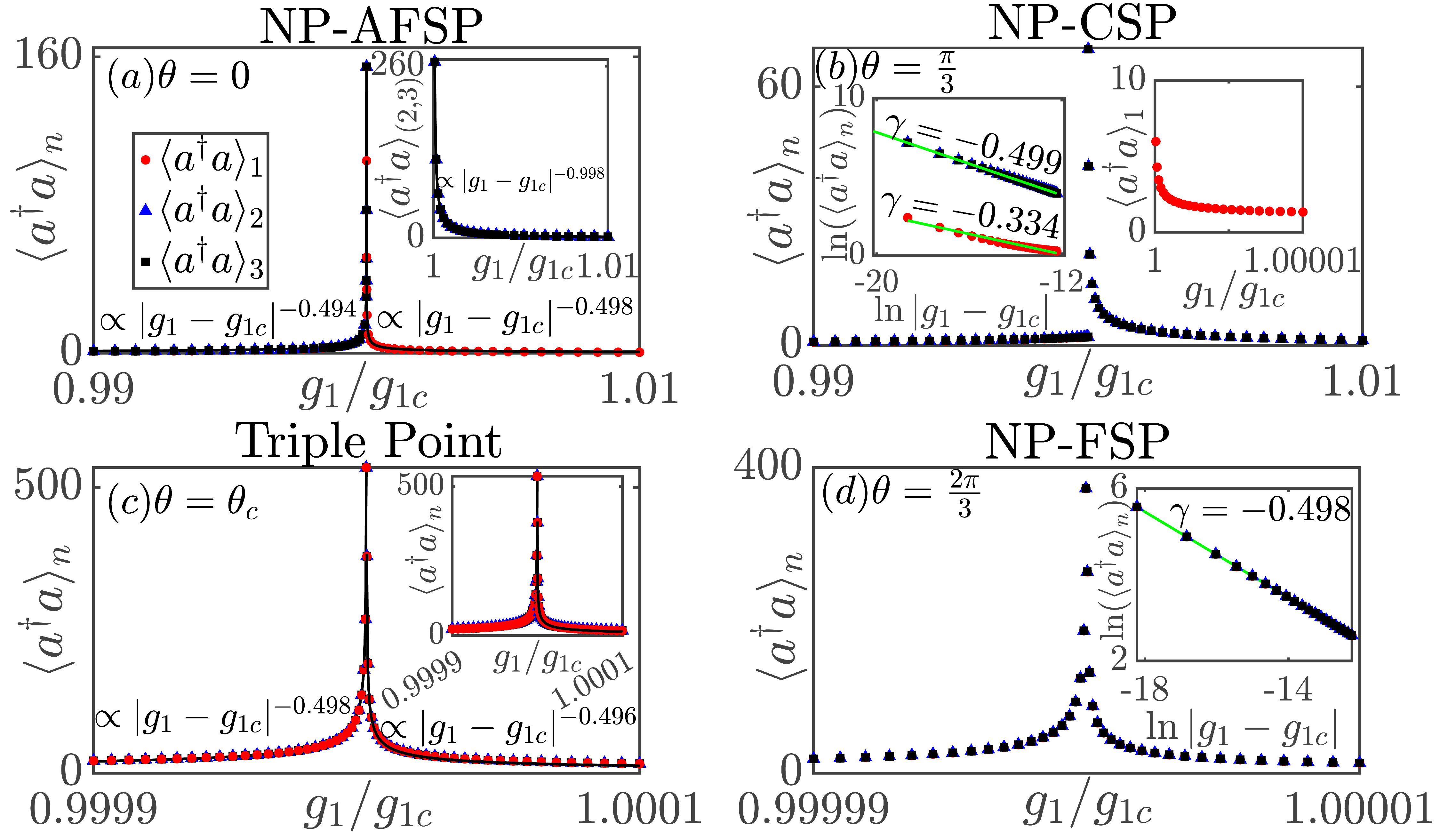}
\caption{Quantum fluctuations of the local mean photons for the $n$th cavity $a^\dagger_na_n$ as a function of $g_1/g_{1c}$ for the NP-AFSP transition ($\theta=0$)(a), the NP-CSP transition ($\theta=0.1<\theta_c$) (b), the triple point ($\theta=\theta_c$) (c), and the NP-FSP transition ($\theta=1.7>\theta_c$) (d). The inset (a) shows the scaling behavior of $a^\dagger_{2(3)}a_{2(3)}$. The right side of the inset (b) shows the behavior of $a^\dagger_{1}a_{1}$, and the left side shows the scaling exponents. The inset (c) shows the scaling results by the solution in the CSP, which is same as one obtained by the solution in the FSP. The inset (d) shows the exponent . }
\label{photons}
\end{figure}

\begin{figure}[tbp]
\includegraphics[scale=0.08]{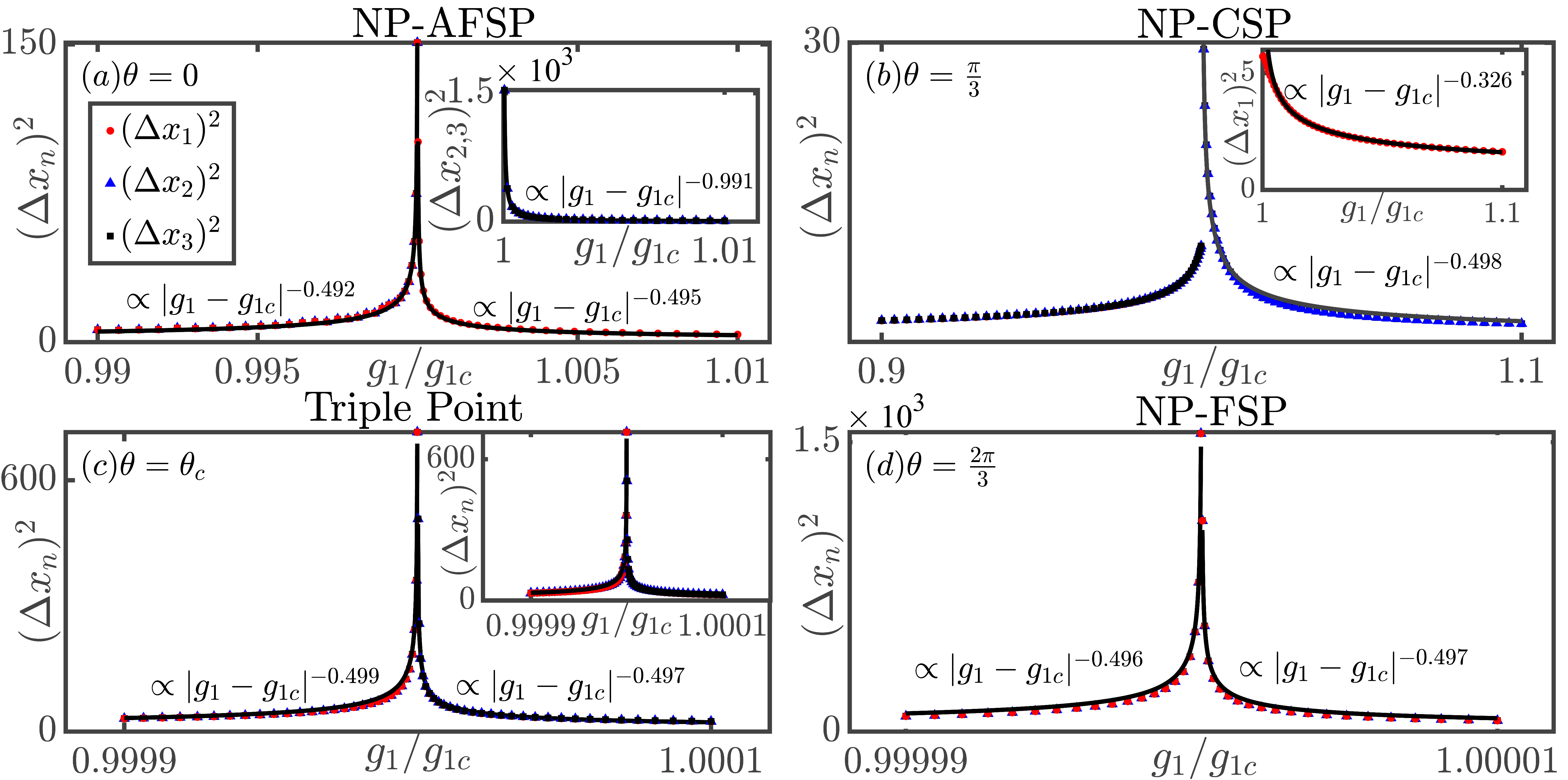}
\caption{Quantum fluctuation of the variance of position for the $n$th cavity $(\Delta x_n)^2$  as a function of $g_1/g_{1c}$ for the NP-AFSP transition ($\theta=0$)(a), the NP-CSP transition ($\theta=0.1<\theta_c$) (b), the triple point ($\theta=\theta_c$) (c), and the NP-FSP transition ($\theta=1.7>\theta_c$) (d). The inset (a) shows the scaling behavior of $(\Delta x_{2(3)})^2$. The inset (b) shows the fluctuations of the first frustrated cavity $(\Delta x_1)^2$ above $g_{1c}$. The inset (c) shows the scaling results by the solution in the CSP, which is same as one obtained by the solution in the FSP.}
\label{fluctuation}
\end{figure}

For the second-order phase transition between the NP and the frustrated AFSP with $\theta=0$, Fig.~\ref{excitation}(a) shows $\varepsilon _{1}$ closing the gap at $g_{1c}$ with the exponent $1/2$ for $g_1<g_{1c}$. In contrast, two excitation energies $\varepsilon _{1}$ and $\varepsilon _{2}$ vanish at the critical point  for $g_1>g_{1c}$
\begin{eqnarray}\label{scal1}
\varepsilon_{1,\texttt{AFSP}}\propto|g_1-g_{1c}|^{-1}, \varepsilon_{2,\texttt{AFSP}}\propto|g_1-g_{1c}|^{-1/2}.
\end{eqnarray} 
It exhibits two different exponents $\gamma_+$ ($\gamma_-$) at the two sides of phase transition for $g_1>g_{1c}$ ($g<g_{1c}$), which is distinguished from the conventional second-order phase transitions. The universal exponent $\gamma_-=1/2$ is the same as a mean-field exponent of the Rabi model~\cite{hwang2015}. While two different exponents $\gamma_{+}=1/2$ and $\gamma_{+}=1$ in Eq.(\ref{scal1}) appear for $g_1>g_{1c}$ , for which the latter is the unconventional scaling exponent beyond the mean-field one. The distinct scaling behavior are consistent with results in the Dicke trimer~\cite{PRLHuang2022}. Meanwhile, Fig.~\ref{photons}(a) shows the fluctuations of the mean photons in each cavity diverging with the same exponent $\beta_-=1/2$ for $g_1<g_{1c}$. However, for $g_1>g_{1c}$, the photon number diverges locally dependent on $n$, which is associated with the frustrated geometry. 
$\langle a_{n}^{\dagger}a_{n}\rangle$ for the $n$th cavity exhibits different scaling law as  
\begin{eqnarray}
\langle a_{1}^{\dagger }a_{1}\rangle_{\texttt{AFSP}}\propto|g_1-g_{1c}|^{-1},\nonumber\\
\langle a_{2(3)}^{\dagger }a_{2(3)}\rangle_{\texttt{AFSP}}\propto|g_1-g_{1c}|^{-1/2}.
\end{eqnarray} 
The scaling of the frustrated cavity ($n=1$) yields an unusual exponent $\beta=1$ , while the remaining cavities ($n=2,3$) diverges with the same exponent $\beta=1/2$ as that in the Rabi model, respectively. The similar $n$-dependent fluctuation of $(\Delta x_n)^2$ is shown in Fig.~\ref{fluctuation}(a). The scaling exponent of $\Delta x_n$ of each cavity below the critical value $g_{1c}$ is the same $\nu_-=1/4$ for $g_1<g_{1c}$. By contrast, for $g_1>g_{1c}$ the fluctuation diverges respectively 
\begin{eqnarray}
(\Delta x_1)^2_{\texttt{AFSP}}\propto|g_1-g_{1c}|^{-1}, \nonumber\\
(\Delta x_{2(3)})^2_{\texttt{AFSP}}\propto|g_1-g_{1c}|^{-1/2}.
\end{eqnarray}
The scaling exponents are extracted $\nu_+=1/2$ for the frustrated cavity and $\nu_+=1/4$ for the remaining cavities. 

For the NP-CSP transition with $0<\theta<\theta_c$, the unusual mode $\varepsilon_1$ closes the gap for $g_1$ approaching $g_{1c}$ from below or above in Fig.~\ref{excitation}(b). It behaves as 
\begin{eqnarray}
\varepsilon_{1,\texttt{NP}}\propto|g_1-g_{1c}|^{-1}.
\varepsilon_{1,\texttt{CSP}}\propto|g_1-g_{1c}|^{-3/2}.
\end{eqnarray}
It gives two unconventional exponents $\gamma_{-}=1$ and $\gamma_+=3/2$ below and above $g_{1c}$, which are different from results in the NP-AFSR transition. In contrast to the divergence at $g_{1c}$, one observes a finite value of $\langle a_{n}^{\dagger }a_{n}\rangle_{\texttt{NP}}$  and $(\Delta x_n)_{\texttt{NP}}$ when $g_1$ approaches $g_{1c}$ from below in Fig.~\ref{photons} (b) and Fig.~\ref{fluctuation}(b). The results are consistent with the analysis from Eq.(\ref{local photon}) and (\ref{position}), for which the denominator is nonzero at the critical point for $q=\pm2\pi/3$. For $g_1>g_{1c}$, Fig.~\ref{photons}(b) shows $n$-dependent scaling behaviors of the local photon number
\begin{eqnarray}
\langle a_{1}^{\dagger }a_{1}\rangle_{\texttt{CSP}}\propto|g_1-g_{1c}|^{-1/3}, \nonumber\\
\langle a_{2(3)}^{\dagger }a_{2(3)}\rangle_{\texttt{CSP}}\propto|g_1-g_{1c}|^{-1/2}.
\end{eqnarray}
$\langle a_{1}^{\dagger }a_{1}\rangle$ of the frustrated cavity diverges with the exponents $1/3$, which is different from that in the AFSP. The anomalous exponent $\nu_+=1/3$ of the frustrated cavity is more accurate by comparing to that in the Dicke lattice~\cite{zhao2022anomalous}. Meanwhile, Fig.~\ref{fluctuation}(b) shows the fluctuation of the position variance of the $n$th cavity, which diverges as 
\begin{eqnarray}
(\Delta x_1)^2_{\texttt{CSP}}\propto|g_1-g_{1c}|^{-1/3}, \nonumber\\
(\Delta x_{2(3)})^2_{\texttt{CSP}}\propto|g_1-g_{1c}|^{-1/2}.
\end{eqnarray}
The unusual exponent $\nu_+=1/6$ of the frustrated cavity is also different from $\nu_+=1/2$ in the AFSP transition. Since the CSR phase transition is associated with both of the geometry frustration and the artificial magnetic field. It exhibits distinct exponents of the frustrated cavity by comparing to the frustrated AFSP transition with $\theta=0$. 
They belong to different universality classes of phase transitions, which are distinguished from the conventional superradiant phase transition in the Dicke and Rabi model.

At the triple point $\theta_c$, the CSR phase, FSP and NP coexist. There appear two excitation modes closing the gaps at $g_{1c}$. Fig.~\ref{excitation}(c) show two different
power laws behaviors
\begin{eqnarray}
\varepsilon_{1,\texttt{TP}}\propto|g_1-g_{1c}|^{-1}, \varepsilon_{2,\texttt{TP}}\propto|g_1-g_{1c}|^{-1/2}.
\end{eqnarray} 
 We obtain exponents $\gamma_{\pm}=1$ and $\gamma_{\pm}=1/2$. It is a signature of the coexistence of both CSP and FSP. And scaling exponents are the same at the two sides of the TP. We calculate the the fluctuation of $\langle a_{n}^{\dagger }a_{n}\rangle$ by the analytical solutions in the FSP phase in Fig.~\ref{photons}(c), which behaves the same as the solution in the CSR in the inset. Fig.~\ref{fluctuation}(c) show the scaling laws of $(\Delta x_n)^2_{\texttt{TP}}$. We obtain the scaling behaviors near the critical point
\begin{eqnarray} 
\langle a_{n}^{\dagger }a_{n}\rangle_{\texttt{TP}}\propto|g_1-g_{1c}|^{-1/2},
(\Delta x_n)^2_{\texttt{TP}}\propto|g_1-g_{1c}|^{-1/2}.
\end{eqnarray}
Thus, we obtain the scaling exponents $\beta=1/2$ and $\nu=1/4$ for the fluctuations of photon number and the position variance at the triple point, respectively.

For the NP-FSP transition, the scaling laws holds  with the same value of the exponent $\gamma$ both for $g_1>g_{1c}$ and $g<g_{1c}$. The exponent of the excitation energy at the two sides of the
phase transition are the same $\gamma_+=\gamma_-=1/2$ in Fig.~\ref{excitation}(d). And the mean photons of the $n$th cavity diverges with the same exponent $\beta_{\pm}=1/2$ in Fig.~\ref{photons}(d). 
Fig.~\ref{photons}(d) show the exponent $\nu_{\pm}=1/4$ for the fluctuation of the position variance. 

Various critical exponents for $g_1$ approaching $g_{1c}$ from below and above are listed in Table I for different superradiant phase transitions. Scaling exponents below and above $g_{1c}$ are different for the NP-AFSP and NP-CSP transition, while are the same for the NP-FSP transition. For the special frustrated cavity, the unusual critical exponent of the photon fluctuations are $\beta_+=1$ and $\beta_+=1/3$  for the NP-AFSP and NP-CSP transition, respectively. Meanwhile, the anomalous exponents of the fluctuations of the position variance are $\nu_+=1/2$ and $\nu_+=1/6$.  They are distinguished from the exponents of the remaining cavities with $\beta_+=1/2$ and $\nu_+=1/4$, which are the same exponents as that in the conventional quantum Rabi model. It demonstrates that the artificial magnetic field and geometric fluctuations yield unusual exponents, which predict nontrivial universality classes of superradiant phase transitions.

\begin{table}[t]\label{exponents}
\caption{ Scaling exponents $\gamma_{\pm}$ of the excitation energy $\varepsilon_n$, $\nu_{\pm}$ and $\beta_{\pm}$ for quantum fluctuations of the $\Delta x_n$ and $\langle a_{n}^{\dagger }a_{n}\rangle$ on two sides of phase transitions of the NP-AFSP($\theta=0$), NP-CSP ($\theta<\theta_c$),  and NP-FSP ($\theta>\theta_c$) as well as the triple point (TP) ($\theta=\theta_c$).}%
\begin{tabularx}{0.49\textwidth} {
  | >{\centering\arraybackslash}X
  | >{\centering\arraybackslash}X
  | >{\centering\arraybackslash}X
  | >{\centering\arraybackslash}X
  | >{\centering\arraybackslash}X
  | >{\centering\arraybackslash}X
  | >{\centering\arraybackslash}X|}
 \hline
    & $\gamma_-$ & $\gamma_+$ & $\nu_-$  & $\nu_{+}$ & $\beta_{-}$ & $\beta_{+}$ \\
  \hline
  NP-AFSP  & $1/2$ & $1,1/2$  & $1/4$  & $1/2,1/4$  & $1/2$ & $1,1/2$   \\
  \hline
  NP-CSP & $1$ & $3/2$ & $/$  & $1/6,1/4$ & $/$ & $ 1/3,1/2$  \\
  \hline
  TP & $1,1/2$ & $1,1/2$  & $1/4$ & $1/4$ & $1/2$ & $1/2$   \\
 \hline
  NP-FSP & $1/2$ & $1/2$ & $1/4$ & $1/4$ &$1/2$ & $1/2$ \\
 \hline
\end{tabularx}
\end{table}

\section{ Conclusion}
we have demonstrated quantum fluctuations and scaling behaviors by using an analytical approach and a Bogoliubov transformation in the Rabi triangle with an artificial magnetic field. The time-reversal symmetry breaking is observed from the photon population dynamics, exhibiting the effects of an artificial magnetic field. Due to the frustrations of the triangular geometry, there emerges two scaling laws, one for the frustrated cavity with an unusual exponent and the other for the remaining cavities with the same exponent as the conventional Dicke model. For the fluctuations of the local photon number and the position variance, the scaling exponents of the  frustrated cavity are $\beta_+=1/3$ and $\nu_+=1/6$ for the CSP transition, which are distinct from $\beta_+=1$ and $\nu_+=1/2$ for the frustrated AFSP transition without an artificial magnetic flux. Moreover, the scaling exponents are different on two sides of the phase transition from the NP to the CSP and AFSP.
The unconventional critical exponents predict different universality classes beyond the conventional single-cavity Rabi model. Therefore, our work paves a way for exploring unconventional phase transitions in the few-body light-matter interacting system.

\begin{acknowledgments}
The authors thank Qing-Hu Chen and Xiang-You Chen for useful discussions.
This work was supported by NSFC under Grant No.12075040 and No. 12347101.
\end{acknowledgments}

\appendix
\section{Derivations of quantum fluctuations}
In the superradiant phase, the operator $a_n$ is shifted as $a_n+\alpha_n$. Using the Bogoliubov's diagonalization method, the ground state of the photon part is obtained as  $|\varphi_{\texttt{SR}}\rangle =|0\rangle_{b_1}|0\rangle_{b_2} |0\rangle_{b_3}$ with $b_n|0\rangle_{b_n}=0$.  With the transformation $D(\alpha)$ and the transformation $\alpha ^{\dagger }=T\beta ^{\dagger }$, the operator $a$ becomes $a_{n}=\sum_{i=1}^{3}T_{n,i}b_{i}+T_{n,i+3}b_{i}^{\dagger}+\alpha_n$. The expected value of $x_{n}=(a_{n}+a_{n}^{\dagger })$ is given by 
\begin{eqnarray}
\langle x_{n}\rangle &=&_{b_n}\langle 0|(a_{n}+a_{n}^{\dagger })|0\rangle_{b_n}=\alpha_n+\alpha_n^* \nonumber\\
&+&_{b_n}\langle 0|\sum_{i=1}^{3}T_{n,i}b_{i}+T_{n,i+3}b_{i}^{\dagger}+T^{*}_{n,i}b_{i}^{\dagger}+T_{n,i+3}^{*}b_{i}|0\rangle_{b_n} \nonumber\\
&=&\alpha _{n}^{\ast }+\alpha _{n}.
\end{eqnarray}
Then we derive the mean value of $x_{n}^{2}$ as
\begin{eqnarray*}
\langle x_{n}^{2}\rangle &=&_{b_n}\langle 0| (a_{n}+a_{n}^{\dagger }+\alpha
_{n}^{\ast }+\alpha _{n})^{2}|0\rangle_{b_n}\\
&=&\langle (a_{n}+a_{n}^{\dagger })^{2}+(\alpha _{n}^{\ast }+\alpha
_{n})(a_{n}+a_{n}^{\dagger })\rangle +(\alpha _{n}^{\ast }+\alpha _{n})^{2}
\\
&=&\langle \varphi_{\texttt{SR}}| \lbrack (T_{n1}+T_{n4}^{\ast })b_{1}+(T_{n2}+T_{n5}^{\ast
})b_{2}\\
&&+(T_{n3}+T_{n6}^{\ast })b_{3}+h.c.]^{2}|\varphi_{\texttt{SR}}\rangle
+(\alpha _{n}^{\ast }+\alpha _{n})^{2} \\
&=&|T_{n1}+T_{n4}^{\ast }|^{2}+|T_{n2}+T_{n5}^{\ast
}|^{2}\\
&&+|T_{n3}+T_{n6}^{\ast }|^{2}+(\alpha _{n}^{\ast }+\alpha _{n})^{2}.
\end{eqnarray*}
The expected value of the variance of $\Delta x_{n}$ is obtained as
\begin{eqnarray}
(\Delta x_{n})^{2}&&=\langle x_{n}^{2}\rangle -\langle x_{n}\rangle
^{2}\notag\\
&&=|T_{n1}+T_{n4}^{\ast }|^{2}+|T_{n2}+T_{n5}^{\ast
}|^{2}+|T_{n3}+T_{n6}^{\ast }|^{2}. \nonumber\\
\end{eqnarray}%
Additionally, the expected value of the momentum quadrature $p_{n}=i(a_{n}^{\dagger }-a_{n})$ is
\begin{equation}
\langle p_{n}\rangle =_{b_n}\langle 0| i(a_{n}^{\dagger }-a_{n}+\alpha _{n}^{\ast
}-\alpha _{n})|0\rangle_{b_n} =i(\alpha _{n}^{\ast }-\alpha _{n}).
\end{equation}%
The expected value of $p_{n}^{2}$ is derived as
\begin{eqnarray}
\langle p_{n}^{2}\rangle &=&_{b_n}\langle 0|  -(a_{n}^{\dagger }-a_{n}+\alpha
_{n}^{\ast }-\alpha _{n})^{2}|0\rangle_{b_n} \notag \\
&=&-\langle (a_{n}^{\dagger }-a_{n})^{2}+(\alpha _{n}^{\ast }-\alpha
_{n})(a_{n}^{\dagger }-a_{n})\rangle -(\alpha _{n}^{\ast }-\alpha _{n})^{2}
\notag \\
&=&-\langle| \lbrack (T_{n4}^{\ast }-T_{n1})b_{1}+(T_{n5}^{\ast
}-T_{n2})b_{2}+(T_{n6}^{\ast }-T_{n3})b_{3}  \notag \\
&&-(T_{n4}-T_{n1}^{\ast })b_{1}^{\dagger }-(T_{n5}-T_{n2}^{\ast
})b_{2}^{\dagger }-(T_{n6}-T_{n3}^{\ast })b_{3}^{\dagger }]^{2}\rangle
\notag \\
&&-(\alpha _{n}^{\ast }-\alpha _{n})^{2}  \notag \\
&=&|T_{n4}^{\ast }-T_{n1}|^{2}+|T_{n5}^{\ast }-T_{n2}|^{2}+|T_{n6}^{\ast
}-T_{n3}|^{2}\notag \\
&&-(\alpha _{n}^{\ast }-\alpha _{n})^{2}.
\end{eqnarray}%
Then one obtains the variance of $\Delta p_{n}$
\begin{eqnarray}
(\Delta p_{n})^{2}&&=\langle p_{n}^{2}\rangle -\langle p_{n}\rangle
^{2}\nonumber\\
&&=|T_{n4}^{\ast }-T_{n1}|^{2}+|T_{n5}^{\ast }-T_{n2}|^{2}+|T_{n6}^{\ast
}-T_{n3}|^{2}.\notag \\
\end{eqnarray}

\bibliography{refs}{}

\begin{thebibliography}{29}%
\makeatletter
\providecommand \@ifxundefined [1]{%
 \@ifx{#1\undefined}
}%
\providecommand \@ifnum [1]{%
 \ifnum #1\expandafter \@firstoftwo
 \else \expandafter \@secondoftwo
 \fi
}%
\providecommand \@ifx [1]{%
 \ifx #1\expandafter \@firstoftwo
 \else \expandafter \@secondoftwo
 \fi
}%
\providecommand \natexlab [1]{#1}%
\providecommand \enquote  [1]{``#1''}%
\providecommand \bibnamefont  [1]{#1}%
\providecommand \bibfnamefont [1]{#1}%
\providecommand \citenamefont [1]{#1}%
\providecommand \href@noop [0]{\@secondoftwo}%
\providecommand \href [0]{\begingroup \@sanitize@url \@href}%
\providecommand \@href[1]{\@@startlink{#1}\@@href}%
\providecommand \@@href[1]{\endgroup#1\@@endlink}%
\providecommand \@sanitize@url [0]{\catcode `\\12\catcode `\$12\catcode
  `\&12\catcode `\#12\catcode `\^12\catcode `\_12\catcode `\%12\relax}%
\providecommand \@@startlink[1]{}%
\providecommand \@@endlink[0]{}%
\providecommand \url  [0]{\begingroup\@sanitize@url \@url }%
\providecommand \@url [1]{\endgroup\@href {#1}{\urlprefix }}%
\providecommand \urlprefix  [0]{URL }%
\providecommand \Eprint [0]{\href }%
\providecommand \doibase [0]{https://doi.org/}%
\providecommand \selectlanguage [0]{\@gobble}%
\providecommand \bibinfo  [0]{\@secondoftwo}%
\providecommand \bibfield  [0]{\@secondoftwo}%
\providecommand \translation [1]{[#1]}%
\providecommand \BibitemOpen [0]{}%
\providecommand \bibitemStop [0]{}%
\providecommand \bibitemNoStop [0]{.\EOS\space}%
\providecommand \EOS [0]{\spacefactor3000\relax}%
\providecommand \BibitemShut  [1]{\csname bibitem#1\endcsname}%
\let\auto@bib@innerbib\@empty
\bibitem [{\citenamefont {Hartmann}\ \emph {et~al.}(2006)\citenamefont
  {Hartmann}, \citenamefont {Brandão},\ and\ \citenamefont
  {Plenio}}]{NaturePhysics2006}%
  \BibitemOpen
  \bibfield  {author} {\bibinfo {author} {\bibfnamefont {M.~J.}\ \bibnamefont
  {Hartmann}}, \bibinfo {author} {\bibfnamefont {F.~G. S.~L.}\ \bibnamefont
  {Brandão}},\ and\ \bibinfo {author} {\bibfnamefont {M.~B.}\ \bibnamefont
  {Plenio}},\ }\href {https://doi.org/10.1038/nphys462} {\bibfield  {journal}
  {\bibinfo  {journal} {Nature Physics}\ }\textbf {\bibinfo {volume} {2}},\
  \bibinfo {pages} {849} (\bibinfo {year} {2006})}\BibitemShut {NoStop}%
\bibitem [{\citenamefont {Rossini}\ and\ \citenamefont
  {Fazio}(2007)}]{PhysRevLett2007}%
  \BibitemOpen
  \bibfield  {author} {\bibinfo {author} {\bibfnamefont {D.}~\bibnamefont
  {Rossini}}\ and\ \bibinfo {author} {\bibfnamefont {R.}~\bibnamefont
  {Fazio}},\ }\href {https://doi.org/10.1103/PhysRevLett.99.186401} {\bibfield
  {journal} {\bibinfo  {journal} {Phys. Rev. Lett.}\ }\textbf {\bibinfo
  {volume} {99}},\ \bibinfo {pages} {186401} (\bibinfo {year}
  {2007})}\BibitemShut {NoStop}%
\bibitem [{\citenamefont {Carusotto}\ \emph {et~al.}(2009)\citenamefont
  {Carusotto}, \citenamefont {Gerace}, \citenamefont {Tureci}, \citenamefont
  {De~Liberato}, \citenamefont {Ciuti},\ and\ \citenamefont
  {Imamo\ifmmode~\check{g}\else \v{g}\fi{}lu}}]{PhysRevLett103}%
  \BibitemOpen
  \bibfield  {author} {\bibinfo {author} {\bibfnamefont {I.}~\bibnamefont
  {Carusotto}}, \bibinfo {author} {\bibfnamefont {D.}~\bibnamefont {Gerace}},
  \bibinfo {author} {\bibfnamefont {H.~E.}\ \bibnamefont {Tureci}}, \bibinfo
  {author} {\bibfnamefont {S.}~\bibnamefont {De~Liberato}}, \bibinfo {author}
  {\bibfnamefont {C.}~\bibnamefont {Ciuti}},\ and\ \bibinfo {author}
  {\bibfnamefont {A.}~\bibnamefont {Imamo\ifmmode~\check{g}\else
  \v{g}\fi{}lu}},\ }\href {https://doi.org/10.1103/PhysRevLett.103.033601}
  {\bibfield  {journal} {\bibinfo  {journal} {Phys. Rev. Lett.}\ }\textbf
  {\bibinfo {volume} {103}},\ \bibinfo {pages} {033601} (\bibinfo {year}
  {2009})}\BibitemShut {NoStop}%
\bibitem [{\citenamefont {Schir\'o}\ \emph {et~al.}(2012)\citenamefont
  {Schir\'o}, \citenamefont {Bordyuh}, \citenamefont {\"Oztop},\ and\
  \citenamefont {T\"ureci}}]{PhysRevLett109}%
  \BibitemOpen
  \bibfield  {author} {\bibinfo {author} {\bibfnamefont {M.}~\bibnamefont
  {Schir\'o}}, \bibinfo {author} {\bibfnamefont {M.}~\bibnamefont {Bordyuh}},
  \bibinfo {author} {\bibfnamefont {B.}~\bibnamefont {\"Oztop}},\ and\ \bibinfo
  {author} {\bibfnamefont {H.~E.}\ \bibnamefont {T\"ureci}},\ }\href
  {https://doi.org/10.1103/PhysRevLett.109.053601} {\bibfield  {journal}
  {\bibinfo  {journal} {Phys. Rev. Lett.}\ }\textbf {\bibinfo {volume} {109}},\
  \bibinfo {pages} {053601} (\bibinfo {year} {2012})}\BibitemShut {NoStop}%
\bibitem [{\citenamefont {Vojta}(2003)}]{voita2003}%
  \BibitemOpen
  \bibfield  {author} {\bibinfo {author} {\bibfnamefont {M.}~\bibnamefont
  {Vojta}},\ }\href@noop {} {\bibfield  {journal} {\bibinfo  {journal} {Rep.
  Prog. Phys.}\ }\textbf {\bibinfo {volume} {66}},\ \bibinfo {pages} {2069}
  (\bibinfo {year} {2003})}\BibitemShut {NoStop}%
\bibitem [{\citenamefont {Sachdev}(2011)}]{sachdev2011}%
  \BibitemOpen
  \bibfield  {author} {\bibinfo {author} {\bibfnamefont {S.}~\bibnamefont
  {Sachdev}},\ }\href {https://doi.org/10.1017/CBO9780511973765} {\emph
  {\bibinfo {title} {Quantum Phase Transitions}}},\ \bibinfo {edition} {2nd}\
  ed.\ (\bibinfo  {publisher} {Cambridge University Press},\ \bibinfo {address}
  {Cambridge},\ \bibinfo {year} {2011})\BibitemShut {NoStop}%
\bibitem [{\citenamefont {Cardy}(1996)}]{scaling1996}%
  \BibitemOpen
  \bibfield  {author} {\bibinfo {author} {\bibfnamefont {J.}~\bibnamefont
  {Cardy}},\ }\href@noop {} {\  (\bibinfo {year} {1996})}\BibitemShut {NoStop}%
\bibitem [{\citenamefont {Sondhi}\ \emph {et~al.}(1997)\citenamefont {Sondhi},
  \citenamefont {Girvin}, \citenamefont {Carini},\ and\ \citenamefont
  {Shahar}}]{RevModPhys1997}%
  \BibitemOpen
  \bibfield  {author} {\bibinfo {author} {\bibfnamefont {S.~L.}\ \bibnamefont
  {Sondhi}}, \bibinfo {author} {\bibfnamefont {S.~M.}\ \bibnamefont {Girvin}},
  \bibinfo {author} {\bibfnamefont {J.~P.}\ \bibnamefont {Carini}},\ and\
  \bibinfo {author} {\bibfnamefont {D.}~\bibnamefont {Shahar}},\ }\href
  {https://doi.org/10.1103/RevModPhys.69.315} {\bibfield  {journal} {\bibinfo
  {journal} {Rev. Mod. Phys.}\ }\textbf {\bibinfo {volume} {69}},\ \bibinfo
  {pages} {315} (\bibinfo {year} {1997})}\BibitemShut {NoStop}%
\bibitem [{\citenamefont {Dicke}(1954)}]{PhysRev1954}%
  \BibitemOpen
  \bibfield  {author} {\bibinfo {author} {\bibfnamefont {R.~H.}\ \bibnamefont
  {Dicke}},\ }\href {https://doi.org/10.1103/PhysRev.93.99} {\bibfield
  {journal} {\bibinfo  {journal} {Phys. Rev.}\ }\textbf {\bibinfo {volume}
  {93}},\ \bibinfo {pages} {99} (\bibinfo {year} {1954})}\BibitemShut {NoStop}%
\bibitem [{\citenamefont {Emary}\ and\ \citenamefont
  {Brandes}(2003)}]{Emary03}%
  \BibitemOpen
  \bibfield  {author} {\bibinfo {author} {\bibfnamefont {C.}~\bibnamefont
  {Emary}}\ and\ \bibinfo {author} {\bibfnamefont {T.}~\bibnamefont
  {Brandes}},\ }\href {https://doi.org/10.1103/PhysRevE.67.066203} {\bibfield
  {journal} {\bibinfo  {journal} {Phys. Rev. E}\ }\textbf {\bibinfo {volume}
  {67}},\ \bibinfo {pages} {066203} (\bibinfo {year} {2003})}\BibitemShut
  {NoStop}%
\bibitem [{\citenamefont {Baumann}\ \emph {et~al.}(2011)\citenamefont
  {Baumann}, \citenamefont {Mottl}, \citenamefont {Brennecke},\ and\
  \citenamefont {Esslinger}}]{PhysRevLett2011}%
  \BibitemOpen
  \bibfield  {author} {\bibinfo {author} {\bibfnamefont {K.}~\bibnamefont
  {Baumann}}, \bibinfo {author} {\bibfnamefont {R.}~\bibnamefont {Mottl}},
  \bibinfo {author} {\bibfnamefont {F.}~\bibnamefont {Brennecke}},\ and\
  \bibinfo {author} {\bibfnamefont {T.}~\bibnamefont {Esslinger}},\ }\href
  {https://doi.org/10.1103/PhysRevLett.107.140402} {\bibfield  {journal}
  {\bibinfo  {journal} {Phys. Rev. Lett.}\ }\textbf {\bibinfo {volume} {107}},\
  \bibinfo {pages} {140402} (\bibinfo {year} {2011})}\BibitemShut {NoStop}%
\bibitem [{\citenamefont {Chen}\ \emph {et~al.}(2008)\citenamefont {Chen},
  \citenamefont {Zhang}, \citenamefont {Liu},\ and\ \citenamefont
  {Wang}}]{PhysRevA2008}%
  \BibitemOpen
  \bibfield  {author} {\bibinfo {author} {\bibfnamefont {Q.-H.}\ \bibnamefont
  {Chen}}, \bibinfo {author} {\bibfnamefont {Y.-Y.}\ \bibnamefont {Zhang}},
  \bibinfo {author} {\bibfnamefont {T.}~\bibnamefont {Liu}},\ and\ \bibinfo
  {author} {\bibfnamefont {K.-L.}\ \bibnamefont {Wang}},\ }\href
  {https://doi.org/10.1103/PhysRevA.78.051801} {\bibfield  {journal} {\bibinfo
  {journal} {Phys. Rev. A}\ }\textbf {\bibinfo {volume} {78}},\ \bibinfo
  {pages} {051801} (\bibinfo {year} {2008})}\BibitemShut {NoStop}%
\bibitem [{\citenamefont {Liu}\ \emph {et~al.}(2009)\citenamefont {Liu},
  \citenamefont {Zhang}, \citenamefont {Chen},\ and\ \citenamefont
  {Wang}}]{PhysRevA2009}%
  \BibitemOpen
  \bibfield  {author} {\bibinfo {author} {\bibfnamefont {T.}~\bibnamefont
  {Liu}}, \bibinfo {author} {\bibfnamefont {Y.-Y.}\ \bibnamefont {Zhang}},
  \bibinfo {author} {\bibfnamefont {Q.-H.}\ \bibnamefont {Chen}},\ and\
  \bibinfo {author} {\bibfnamefont {K.-L.}\ \bibnamefont {Wang}},\ }\href
  {https://doi.org/10.1103/PhysRevA.80.023810} {\bibfield  {journal} {\bibinfo
  {journal} {Phys. Rev. A}\ }\textbf {\bibinfo {volume} {80}},\ \bibinfo
  {pages} {023810} (\bibinfo {year} {2009})}\BibitemShut {NoStop}%
\bibitem [{\citenamefont {Hwang}\ \emph {et~al.}(2015)\citenamefont {Hwang},
  \citenamefont {Puebla},\ and\ \citenamefont {Plenio}}]{hwang2015}%
  \BibitemOpen
  \bibfield  {author} {\bibinfo {author} {\bibfnamefont {M.-J.}\ \bibnamefont
  {Hwang}}, \bibinfo {author} {\bibfnamefont {R.}~\bibnamefont {Puebla}},\ and\
  \bibinfo {author} {\bibfnamefont {M.~B.}\ \bibnamefont {Plenio}},\
  }\href@noop {} {\bibfield  {journal} {\bibinfo  {journal} {Phys. Rev. Lett.}\
  }\textbf {\bibinfo {volume} {115}},\ \bibinfo {pages} {180404} (\bibinfo
  {year} {2015})}\BibitemShut {NoStop}%
\bibitem [{\citenamefont {Ashhab}(2013)}]{Ashhab2013}%
  \BibitemOpen
  \bibfield  {author} {\bibinfo {author} {\bibfnamefont {S.}~\bibnamefont
  {Ashhab}},\ }\href@noop {} {\bibfield  {journal} {\bibinfo  {journal} {Phys.
  Rev. A}\ }\textbf {\bibinfo {volume} {87}},\ \bibinfo {pages} {013826}
  (\bibinfo {year} {2013})}\BibitemShut {NoStop}%
\bibitem [{\citenamefont {Liu}\ \emph {et~al.}(2017)\citenamefont {Liu},
  \citenamefont {Chesi}, \citenamefont {Ying}, \citenamefont {Chen},
  \citenamefont {Luo},\ and\ \citenamefont {Lin}}]{liu2017}%
  \BibitemOpen
  \bibfield  {author} {\bibinfo {author} {\bibfnamefont {M.}~\bibnamefont
  {Liu}}, \bibinfo {author} {\bibfnamefont {S.}~\bibnamefont {Chesi}}, \bibinfo
  {author} {\bibfnamefont {Z.-J.}\ \bibnamefont {Ying}}, \bibinfo {author}
  {\bibfnamefont {X.}~\bibnamefont {Chen}}, \bibinfo {author} {\bibfnamefont
  {H.-G.}\ \bibnamefont {Luo}},\ and\ \bibinfo {author} {\bibfnamefont {H.-Q.}\
  \bibnamefont {Lin}},\ }\href@noop {} {\bibfield  {journal} {\bibinfo
  {journal} {Phys. Rev. Lett.}\ }\textbf {\bibinfo {volume} {119}},\ \bibinfo
  {pages} {220601} (\bibinfo {year} {2017})}\BibitemShut {NoStop}%
\bibitem [{\citenamefont {Chen}\ \emph {et~al.}(2020)\citenamefont {Chen},
  \citenamefont {Zhang}, \citenamefont {Fu},\ and\ \citenamefont
  {Zheng}}]{chen2020}%
  \BibitemOpen
  \bibfield  {author} {\bibinfo {author} {\bibfnamefont {X.-Y.}\ \bibnamefont
  {Chen}}, \bibinfo {author} {\bibfnamefont {Y.-Y.}\ \bibnamefont {Zhang}},
  \bibinfo {author} {\bibfnamefont {L.}~\bibnamefont {Fu}},\ and\ \bibinfo
  {author} {\bibfnamefont {H.}~\bibnamefont {Zheng}},\ }\href@noop {}
  {\bibfield  {journal} {\bibinfo  {journal} {Phys. Rev. A}\ }\textbf {\bibinfo
  {volume} {101}},\ \bibinfo {pages} {033827} (\bibinfo {year}
  {2020})}\BibitemShut {NoStop}%
\bibitem [{\citenamefont {Chen}\ \emph {et~al.}(2021)\citenamefont {Chen},
  \citenamefont {Wu}, \citenamefont {Jiang}, \citenamefont {L{\"u}},
  \citenamefont {Peng},\ and\ \citenamefont {Du}}]{chen2021}%
  \BibitemOpen
  \bibfield  {author} {\bibinfo {author} {\bibfnamefont {X.}~\bibnamefont
  {Chen}}, \bibinfo {author} {\bibfnamefont {Z.}~\bibnamefont {Wu}}, \bibinfo
  {author} {\bibfnamefont {M.}~\bibnamefont {Jiang}}, \bibinfo {author}
  {\bibfnamefont {X.-Y.}\ \bibnamefont {L{\"u}}}, \bibinfo {author}
  {\bibfnamefont {X.}~\bibnamefont {Peng}},\ and\ \bibinfo {author}
  {\bibfnamefont {J.}~\bibnamefont {Du}},\ }\href@noop {} {\bibfield  {journal}
  {\bibinfo  {journal} {Nat. Commun.}\ }\textbf {\bibinfo {volume} {12}},\
  \bibinfo {pages} {1} (\bibinfo {year} {2021})}\BibitemShut {NoStop}%
\bibitem [{\citenamefont {Cai}\ and\ \citenamefont {et~al.}(2021)}]{NCcai2021}%
  \BibitemOpen
  \bibfield  {author} {\bibinfo {author} {\bibfnamefont {M.~L.}\ \bibnamefont
  {Cai}}\ and\ \bibinfo {author} {\bibnamefont {et~al.}},\ }\href@noop {}
  {\bibfield  {journal} {\bibinfo  {journal} {Nat. Commun.}\ }\textbf {\bibinfo
  {volume} {12}},\ \bibinfo {pages} {1126} (\bibinfo {year}
  {2021})}\BibitemShut {NoStop}%
\bibitem [{\citenamefont {Dalibard}\ \emph {et~al.}(2011)\citenamefont
  {Dalibard}, \citenamefont {Gerbier}, \citenamefont {Juzeli{\=u}nas},\ and\
  \citenamefont {{\"O}hberg}}]{dalibard2011}%
  \BibitemOpen
  \bibfield  {author} {\bibinfo {author} {\bibfnamefont {J.}~\bibnamefont
  {Dalibard}}, \bibinfo {author} {\bibfnamefont {F.}~\bibnamefont {Gerbier}},
  \bibinfo {author} {\bibfnamefont {G.}~\bibnamefont {Juzeli{\=u}nas}},\ and\
  \bibinfo {author} {\bibfnamefont {P.}~\bibnamefont {{\"O}hberg}},\
  }\href@noop {} {\bibfield  {journal} {\bibinfo  {journal} {Rev. Mod. Phys.}\
  }\textbf {\bibinfo {volume} {83}},\ \bibinfo {pages} {1523} (\bibinfo {year}
  {2011})}\BibitemShut {NoStop}%
\bibitem [{\citenamefont {Cai}\ \emph {et~al.}(2019)\citenamefont {Cai},
  \citenamefont {Liu}, \citenamefont {Wu}, \citenamefont {He}, \citenamefont
  {Zhu}, \citenamefont {Zhang},\ and\ \citenamefont {Wang}}]{PhysRevLett122}%
  \BibitemOpen
  \bibfield  {author} {\bibinfo {author} {\bibfnamefont {H.}~\bibnamefont
  {Cai}}, \bibinfo {author} {\bibfnamefont {J.}~\bibnamefont {Liu}}, \bibinfo
  {author} {\bibfnamefont {J.}~\bibnamefont {Wu}}, \bibinfo {author}
  {\bibfnamefont {Y.}~\bibnamefont {He}}, \bibinfo {author} {\bibfnamefont
  {S.-Y.}\ \bibnamefont {Zhu}}, \bibinfo {author} {\bibfnamefont {J.-X.}\
  \bibnamefont {Zhang}},\ and\ \bibinfo {author} {\bibfnamefont {D.-W.}\
  \bibnamefont {Wang}},\ }\href
  {https://doi.org/10.1103/PhysRevLett.122.023601} {\bibfield  {journal}
  {\bibinfo  {journal} {Phys. Rev. Lett.}\ }\textbf {\bibinfo {volume} {122}},\
  \bibinfo {pages} {023601} (\bibinfo {year} {2019})}\BibitemShut {NoStop}%
\bibitem [{\citenamefont {Li}\ \emph {et~al.}(2020)\citenamefont {Li},
  \citenamefont {Cai}, \citenamefont {Wang}, \citenamefont {Li}, \citenamefont
  {Yuan},\ and\ \citenamefont {Li}}]{PhysRevLett124}%
  \BibitemOpen
  \bibfield  {author} {\bibinfo {author} {\bibfnamefont {Y.}~\bibnamefont
  {Li}}, \bibinfo {author} {\bibfnamefont {H.}~\bibnamefont {Cai}}, \bibinfo
  {author} {\bibfnamefont {D.-w.}\ \bibnamefont {Wang}}, \bibinfo {author}
  {\bibfnamefont {L.}~\bibnamefont {Li}}, \bibinfo {author} {\bibfnamefont
  {J.}~\bibnamefont {Yuan}},\ and\ \bibinfo {author} {\bibfnamefont
  {W.}~\bibnamefont {Li}},\ }\href
  {https://doi.org/10.1103/PhysRevLett.124.140401} {\bibfield  {journal}
  {\bibinfo  {journal} {Phys. Rev. Lett.}\ }\textbf {\bibinfo {volume} {124}},\
  \bibinfo {pages} {140401} (\bibinfo {year} {2020})}\BibitemShut {NoStop}%
\bibitem [{\citenamefont {Roushan}\ \emph {et~al.}(2017)\citenamefont
  {Roushan}, \citenamefont {Neill}, \citenamefont {Megrant}, \citenamefont
  {Chen}, \citenamefont {Babbush}, \citenamefont {Barends}, \citenamefont
  {Campbell}, \citenamefont {Chen}, \citenamefont {Chiaro}, \citenamefont
  {Dunsworth} \emph {et~al.}}]{roushan2017}%
  \BibitemOpen
  \bibfield  {author} {\bibinfo {author} {\bibfnamefont {P.}~\bibnamefont
  {Roushan}}, \bibinfo {author} {\bibfnamefont {C.}~\bibnamefont {Neill}},
  \bibinfo {author} {\bibfnamefont {A.}~\bibnamefont {Megrant}}, \bibinfo
  {author} {\bibfnamefont {Y.}~\bibnamefont {Chen}}, \bibinfo {author}
  {\bibfnamefont {R.}~\bibnamefont {Babbush}}, \bibinfo {author} {\bibfnamefont
  {R.}~\bibnamefont {Barends}}, \bibinfo {author} {\bibfnamefont
  {B.}~\bibnamefont {Campbell}}, \bibinfo {author} {\bibfnamefont
  {Z.}~\bibnamefont {Chen}}, \bibinfo {author} {\bibfnamefont {B.}~\bibnamefont
  {Chiaro}}, \bibinfo {author} {\bibfnamefont {A.}~\bibnamefont {Dunsworth}},
  \emph {et~al.},\ }\href@noop {} {\bibfield  {journal} {\bibinfo  {journal}
  {Nat. Phys.}\ }\textbf {\bibinfo {volume} {13}},\ \bibinfo {pages} {146}
  (\bibinfo {year} {2017})}\BibitemShut {NoStop}%
\bibitem [{\citenamefont {Hayward}\ \emph {et~al.}(2012)\citenamefont
  {Hayward}, \citenamefont {Martin},\ and\ \citenamefont
  {Greentree}}]{hayward2012}%
  \BibitemOpen
  \bibfield  {author} {\bibinfo {author} {\bibfnamefont {A.~L.}\ \bibnamefont
  {Hayward}}, \bibinfo {author} {\bibfnamefont {A.~M.}\ \bibnamefont
  {Martin}},\ and\ \bibinfo {author} {\bibfnamefont {A.~D.}\ \bibnamefont
  {Greentree}},\ }\href@noop {} {\bibfield  {journal} {\bibinfo  {journal}
  {Phys. Rev. Lett.}\ }\textbf {\bibinfo {volume} {108}},\ \bibinfo {pages}
  {223602} (\bibinfo {year} {2012})}\BibitemShut {NoStop}%
\bibitem [{\citenamefont {Hayward}\ and\ \citenamefont
  {Martin}(2016)}]{hayward2016}%
  \BibitemOpen
  \bibfield  {author} {\bibinfo {author} {\bibfnamefont {A.~L.}\ \bibnamefont
  {Hayward}}\ and\ \bibinfo {author} {\bibfnamefont {A.~M.}\ \bibnamefont
  {Martin}},\ }\href@noop {} {\bibfield  {journal} {\bibinfo  {journal} {Phys.
  Rev. A}\ }\textbf {\bibinfo {volume} {93}},\ \bibinfo {pages} {023828}
  (\bibinfo {year} {2016})}\BibitemShut {NoStop}%
\bibitem [{\citenamefont {Zhang}\ \emph {et~al.}(2021)\citenamefont {Zhang},
  \citenamefont {Hu}, \citenamefont {Fu}, \citenamefont {Luo}, \citenamefont
  {Pu},\ and\ \citenamefont {Zhang}}]{PhysRevLettzhang2021}%
  \BibitemOpen
  \bibfield  {author} {\bibinfo {author} {\bibfnamefont {Y.-Y.}\ \bibnamefont
  {Zhang}}, \bibinfo {author} {\bibfnamefont {Z.-X.}\ \bibnamefont {Hu}},
  \bibinfo {author} {\bibfnamefont {L.}~\bibnamefont {Fu}}, \bibinfo {author}
  {\bibfnamefont {H.-G.}\ \bibnamefont {Luo}}, \bibinfo {author} {\bibfnamefont
  {H.}~\bibnamefont {Pu}},\ and\ \bibinfo {author} {\bibfnamefont {X.-F.}\
  \bibnamefont {Zhang}},\ }\href
  {https://doi.org/10.1103/PhysRevLett.127.063602} {\bibfield  {journal}
  {\bibinfo  {journal} {Phys. Rev. Lett.}\ }\textbf {\bibinfo {volume} {127}},\
  \bibinfo {pages} {063602} (\bibinfo {year} {2021})}\BibitemShut {NoStop}%
\bibitem [{\citenamefont {Fallas~Padilla}\ \emph {et~al.}(2022)\citenamefont
  {Fallas~Padilla}, \citenamefont {Pu}, \citenamefont {Cheng},\ and\
  \citenamefont {Zhang}}]{PhysRevLettzhang2022}%
  \BibitemOpen
  \bibfield  {author} {\bibinfo {author} {\bibfnamefont {D.}~\bibnamefont
  {Fallas~Padilla}}, \bibinfo {author} {\bibfnamefont {H.}~\bibnamefont {Pu}},
  \bibinfo {author} {\bibfnamefont {G.-J.}\ \bibnamefont {Cheng}},\ and\
  \bibinfo {author} {\bibfnamefont {Y.-Y.}\ \bibnamefont {Zhang}},\ }\href
  {https://doi.org/10.1103/PhysRevLett.129.183602} {\bibfield  {journal}
  {\bibinfo  {journal} {Phys. Rev. Lett.}\ }\textbf {\bibinfo {volume} {129}},\
  \bibinfo {pages} {183602} (\bibinfo {year} {2022})}\BibitemShut {NoStop}%
\bibitem [{\citenamefont {Zhao}\ and\ \citenamefont
  {Hwang}(2023)}]{zhao2022anomalous}%
  \BibitemOpen
  \bibfield  {author} {\bibinfo {author} {\bibfnamefont {J.}~\bibnamefont
  {Zhao}}\ and\ \bibinfo {author} {\bibfnamefont {M.-J.}\ \bibnamefont
  {Hwang}},\ }\href {https://doi.org/10.1103/PhysRevResearch.5.L042016}
  {\bibfield  {journal} {\bibinfo  {journal} {Phys. Rev. Res.}\ }\textbf
  {\bibinfo {volume} {5}},\ \bibinfo {pages} {L042016} (\bibinfo {year}
  {2023})}\BibitemShut {NoStop}%
\bibitem [{\citenamefont {Zhao}\ and\ \citenamefont
  {Hwang}(2022)}]{PRLHuang2022}%
  \BibitemOpen
  \bibfield  {author} {\bibinfo {author} {\bibfnamefont {J.}~\bibnamefont
  {Zhao}}\ and\ \bibinfo {author} {\bibfnamefont {M.-J.}\ \bibnamefont
  {Hwang}},\ }\href {https://doi.org/10.1103/PhysRevLett.128.163601} {\bibfield
   {journal} {\bibinfo  {journal} {Phys. Rev. Lett.}\ }\textbf {\bibinfo
  {volume} {128}},\ \bibinfo {pages} {163601} (\bibinfo {year}
  {2022})}\BibitemShut {NoStop}%
\end{thebibliography}%

\end{document}